\documentclass[twocolumn,english,showpacs,preprintnumbers,amsmath,amssymb,prb]{revtex4-1}
\usepackage{color}
\usepackage{graphicx}

\makeatletter

\providecommand{\tabularnewline}{\\}

\usepackage{epsfig}

\makeatother

\usepackage{babel}
\begin{document}

\title{Anharmonicity of the acoustic modes of graphene}

\author{R. Ram\'{\i}rez\footnote{Electronic mail:ramirez@icmm.csic.es} and
C. P. Herrero}

\affiliation{Instituto de Ciencia de Materiales de Madrid (ICMM), Consejo Superior
de Investigaciones Cient\'{\i}ficas (CSIC), Campus de Cantoblanco,
28049 Madrid, Spain }
\begin{abstract}
The anharmonicity of the acoustic phonon dispersion of graphene has
been studied by the harmonic linear response (HLR) approach at finite
temperature. This is a non-perturbative method based on the linear
response of the system to applied forces, as derived from equilibrium
computer simulations. Anharmonic shifts are analyzed in the long-wavelength
limit at room temperature, with emphasis in the effect of applied
tensile or compressive in-plane stress. The simulation results are
compared with available analytical models, based either on first-order
perturbation theory or on a description by anomalous exponents. The
simulations show better agreement to the expectations of the perturbational
approach. The effect of temperature and zero-point vibrations on the
acoustic out-of-plane anharmonic shifts of graphene are briefly reviewed. 
\end{abstract}
\maketitle

\section{Introduction}

Geim and Novoselov made the seminal discovering that graphene, a two
dimensional, one-atom width layer found as stacks in graphite, can
be manipulated as a stable planar layer.\citep{novoselov04} Graphene
flakes have been characterized by optical, electrical, mechanical,
and transport methods.\citep{amorim16,cooper_12} Two-dimensional
(2D) crystalline solids were expected to loose their long-range ordering,
as a consequence of spatial atomic fluctuations caused by zero-point
vibrations and temperature.\citep{mermin66} The unexpected stability
of crystalline flat membranes, like graphene, has been explained by
\textit{anharmonic effects}, caused by the coupling between in-plane
and out-of-plane vibrational modes.\citep{Los09,amorim14}

The simplest treatment of anharmonic effects in solids is the quasi-harmonic
approximation (QHA). This approach considers that the angular frequency,
$\omega,$ of each phonon may change with the volume of the crystal
(or in-plane area in 2D solids) but does not depend explicitly on
temperature. The volume dependence is described by the Gruneisen constants,
$\gamma_{\omega}$, and the harmonic limit implies $\gamma_{\omega}=0$.
The usual quasi-harmonic behavior corresponds to values $\gamma_{\omega}>0$,
but certain solids display $\gamma_{\omega}<0$ for some phonons.
Negative Gruneisen constants appear in solids with tetrahedral coordination,
as diamond, silicon, quartz, or ice Ih, and also in anisotropic structures
such as graphite and graphene.\citep{pamuk_12,mounet05} A set of
long-wavelength acoustic modes with $\gamma_{\omega}<0$ is the origin
of the negative thermal expansion of solids at low temperatures. The
negative thermal expansion of graphene has been studied both experimentally
and theoretically by the QHA and perturbation theory methods, as well
as by computer simulations.\citep{storch_18,mounet05,herrero18,michel15}
The shortcomings of the QHA to describe anharmonic effects in graphene
is evident by its incapability to reproduce the crossover from a negative
to a positive thermal expansion coefficient as the temperature increases.\citep{mounet05} 

Anharmonic effects beyond those described by the QHA are often called
\textit{explicitly anharmonic.\citep{grimvall_99} }They are caused
by higher-than-quadratic terms in the potential energy. They produce
frequency shifts, with respect to the harmonic limit, that depend
explicitly on the temperature. The explicit anharmonicity of the ZA
out-of-plane acoustic modes of graphene has been analyzed by using
perturbation theory.\citep{amorim14,michel15prb,michel15,adamyan16,bondarev18}
The main perturbational result is that, in the long-wavelength limit,
the harmonic (H) acoustic ZA dispersion $\rho\omega_{ZA}^{2}=\kappa k^{4}$
of an \textit{unstressed} graphene layer is renormalized to: 
$\rho\omega_{ZA}^{2}=\sigma k^{2}+\kappa k^{4}$.
Here $\rho$ is the surface density, $\kappa$ is the bending rigidity,
$\sigma$ is the fluctuation tension, and $k$ is the modulus of the
wavevector $\mathbf{k}$. The tension $\sigma$ defines the lowest-order
wavevector dependence of the spatial out-of-plane 
fluctuations.\citep{fournier08,shiba_16}

Two physical consequences are derived from the explicit anharmonicity
of the ZA phonons in graphene. The first is related to the mean quadratic
amplitude, $h^{2}$, of the out-of-plane modes. In the \textit{harmonic
limit, }the quadratic amplitude increases linearly with the number
of layer atoms, $h^{2}\sim N,$ leading to a catastrophic divergence
in the thermodynamic limit. However, the anharmonic renormalization
displays a less critical divergence as $h^{2}\sim\ln N$, explaining
an increased stabilization of the flat layer. The second consequence
is that the acoustic sound velocity $v_{ZA}=(\partial\omega_{ZA}/\partial k)_{k=0}$
vanishes in the harmonic limit, but becomes finite by explicit anharmonicity.\citep{adamyan16,bondarev18}

An alternative theoretical framework to explain the stability of a
flat graphene layer is based on the description of the explicit anharmonicity
by a $k-$dependent renormalization of the bending constant, $\kappa,$
giving rise to an acoustic ZA dispersion $\rho\omega_{ZA}^{2}=\kappa k^{4-\eta}$.
$\eta$ is a positive \textit{anomalous exponent} that was estimated
as $\eta=0.82$ within the self-consistent screening approximation.\citep{doussal17}
The numerical analysis of out-of-plane amplitudes in several computer
simulations of graphene provides values in a range, $\eta=0.67-1.1$.
\citep{gao14,Los09} However, the exponent $\eta$ is believed to
be a universal quantity, so that the variability reported in computer
simulation remains unexplained.\citep{amorim16} Within this model,
the mean quadratic amplitude, $h^{2}\sim N^{1-(\eta/2)}$, becomes
much smaller than the harmonic expectation, stabilizing the flat layer.
However, the acoustic sound velocity $v_{ZA}$ vanishes, in disagreement
to the perturbation theory results.

The question about which model (perturbation theory vs anomalous exponent)
provides better agreement to experiment has not been unambiguously
answered, as neither the dispersion relation of the ZA modes in the
long-wavelength limit nor the associated sound velocity have been
measured yet. Several computer simulations published so far are interpreted
in terms of the anomalous exponent model,\citep{Los09,roldan11,gao14,los16a,hasik18}
but there are exceptions.\citep{ramirez16} In spite of the absence
of definite evidence, there seems to be a certain consensus that the
anomalous exponent model is the correct one for graphene.\citep{amorim16}

In this paper we study the explicit anharmonicity of the acoustic
phonon dispersion bands of graphene (two in-plane and one ZA branches)
using the empirical long-range carbon bond order (LCBOPII) model.\citep{los05}
The harmonic linear response (HLR) method is a non-perturbative approach
used to study \textit{anharmonic} vibrations from the analysis of
spatial trajectories generated by equilibrium simulations. The method
was originally proposed in the framework of quantum path-integral
(PI) simulations.\citep{ramirez01} It has been recently applied to
derive the dispersion bands of 2D solids such as a graphene monolayer,
a graphene bilayer, and graphane. \citep{ramirez_19} The explicit
anharmonicity derived by the HLR method will be compared to the expectations
of the available analytical models (perturbation theory vs anomalous
exponent).

The paper is organized as follows. The computational method for the
calculation of the phonon dispersion of graphene is presented in Sec.
\ref{sec:Computational-conditions}. The expectation of the QHA is
discussed in Sec. \ref{sec:QHA-approach}. The study of the explicit
anharmonicity of the acoustic bands of graphene at 300 K as a function
of the in-plane stress is found in Sec. \ref{sec:Explicit-anharmonicty}.
The comparison of simulation results and analytical models is the
focus of Sec. \ref{sec:Comparison-to-analytical}. Temperature and
quantum effects in the anharmonicity of the out-of-plane modes of
graphene are commented in Sec. \ref{sec:Temperature-and-quantum}.
The paper closes with a summary. 

\section{Computational Method\label{sec:Computational-conditions} }

In this section, a minimum set of technical details is presented concerning
the simulation method and the calculation of phonon dispersion relations
in graphene. Further technical information can be found in Refs. \onlinecite{ramirez16,ramirez_19}. 

\subsection{MD simulations}

Classical molecular dynamics (MD) simulations of
graphene were performed in the $N\tau T$ ensemble at temperature
$T=300$ K and applied in-plane stress, $\tau,$ between -0.02 and
0.01 eV/$\textrm{Å}^{2}$. $\tau>0$ ($\tau<0$ ) implies compressive
(tensile) stress. 
The interatomic forces were calculated
with a realistic interatomic potential, namely, the so-called LCBOPII
model. This is a long-range carbon bond order potential, fitted to
ab-initio electronic structure calculations, aiming at the description
of carbon liquid and solid phases, first of all graphite and diamond,
as accurately as possible.\citep{los05} It has been employed earlier
to perform classical simulations of liquid carbon,\citep{ghiringhelli_05b}
diamond, graphite,\citep{los05} and graphene layers\citep{zakharchenko_09,fasolino07,los16a}.
It has been used to predict the carbon phase diagram comprising graphite,
diamond and the liquid, showing that the graphite-diamond transition
line is in good agreement with experimental data.\citep{ghiringhelli05}
The LCBOPII model has been found to give a good description of the
elastic properties, such as the Young's modulus of 
graphene.\citep{zakharchenko_09,politano12}
A brief account of the empirical LCBOPII model is presented in Appendix
\ref{sec:LCBOPII-Potential}.

The simulation cell was a rectangular one with $N=8400$
carbon atoms and similar lengths ($L\sim148\;\textrm{Å}$) along the
$x-$ and $y-$axis in the plane of the layer. The in-plane area per
atom is denoted as $A_{p}.$ Periodic boundary conditions were applied
to the simulation cell in the $xy-$plane. The MD simulations of graphene
were performed in the $N\tau T$ ensemble by allowing isotropic fluctuations
of the in-plane area. The equations of motion, which are summarized
in Appendix \ref{sec:Dynamic-Equations-for}, were integrated using
different time steps for the fast and slow degrees of freedom.\citep{ma96}
The employed time step for the calculation of interatomic
forces was $\triangle t=1$ fs. For the time evolution of the thermostats
and barostat variables we used a time step of $\triangle t/4$, as
in earlier simulations.\citep{herrero14} The equilibration run comprised
$10^{5}$ MD steps. Trajectories with $S=5\times10^{4}$ spatial configurations
were stored for further analysis at equidistant intervals from a long
simulation run with $10^{7}$ MD steps. Long trajectories are mandatory
for a reasonable sampling of the sluggish long-wavelengths acoustic
modes.

\subsection{Phonon dispersion calculation}

The HLR method is a non-perturbative approach to obtain the phonon
dispersion relations from the analysis of the spontaneous atomic fluctuations
of the system, by either classical or quantum PI simulations. This
method requires only spatial information (i.e., not atomic velocities).
This is an advantage in PIMD simulations as the atomic velocities
do not carry true physical information on the dynamics of the quantum
particles.\citep{ramirez02} The physical basis and applicability
of this method to 2D solids has been explained in detail in Ref. \onlinecite{ramirez_19}.
For this reason, only a succinct sketch of the method is given here. 

From the stored trajectory one needs to calculate the in-plane equilibrium
positions of the cell atoms, $\mathbf{r}_{eq,\alpha j}$. $\alpha$
is an index (1 or 2) that runs over the 2 basis atoms of a primitive
cell, while $j$ is an index running over all the basis atoms $\alpha$
in the simulation cell ($N/2$ $\alpha-$type atoms). The equilibrium
$z-$coordinate of the atoms in the flat layer can be set as $z_{eq}=0,$
without loss of generality. 

If the instantaneous displacement vector of an atom from its average
position ($\mathbf{r}_{eq,\alpha j},z_{eq}$) is denoted as ($X_{\alpha j}$,$Y_{\alpha j}$,$Z_{\alpha j}$),
one needs to calculate symmetry adapted Bloch functions, $\overline{X}_{\alpha}(\mathbf{k})$,
with the displacement coordinates as
\begin{equation}
\overline{X}_{\alpha}(\mathbf{k})=\sqrt{\frac{2m}{N}}\sum_{j=1}^{N/2}X_{\alpha j}\exp\left(i\mathbf{k}\mathbf{r}_{eq,\alpha j}\right)\:,\label{eq:FT}
\end{equation}
where $m$ is the carbon mass, and $\mathbf{k}$ is a wavevector commensurate
with the employed simulation cell.\citep{ramirez_19} The number of
Bloch functions is $6,$ i.e., $\left[\overline{X}_{1}(\mathbf{k}),\overline{Y}_{1}(\mathbf{k}),\overline{Z}_{1}(\mathbf{k}),\overline{X}_{2}(\mathbf{k}),\overline{Y}_{2}(\mathbf{k}),\overline{Z}_{2}(\mathbf{k})\right]$,
which corresponds to the number of vibrational bands in graphene.
The covariance of symmetry adapted displacement coordinates is calculated as:
\begin{equation}
\left\langle \overline{C}_{\alpha}(\mathbf{k})\overline{D}_{\beta}^{*}(\mathbf{k})\right\rangle =S^{-1}\sum_{s=1}^{S}\left(\overline{C}_{\alpha}(\mathbf{k})\overline{D}_{\beta}^{*}(\mathbf{k})\right)_{s}\:,
\end{equation}
 where $s$ is a running index for the stored trajectories, $C$ and
$D$ are any of the coordinates $(X,Y,Z)$, and $\alpha$ and $\beta$
are any of the basis atoms $(1,2)$. The covariances form a $6\times6$
tensor, $\chi(\mathbf{k})$. The $j$th eigenvalue, $\varDelta_{j}(\mathbf{k})$,
of the tensor $\chi(\mathbf{k})$ provides an estimation of the angular
frequency associated to the $j$th phonon branch of the 2D layer as,\citep{ramirez_19}
\begin{equation}
\omega_{j}^{2}(\mathbf{k})=\frac{k_{B}T}{\varDelta_{j}(\mathbf{k})}\:,
\end{equation}
where $k_{B}$ is the Boltzmann constant. Within the grid of wavevectors
$\mathbf{k}$ commensurate with the simulation cell, the two $\mathbf{k}$-vectors
with largest wavelength, oriented along either the $x-$ or $y-$directions,
will be denoted as $\mathbf{k}_{min}$. The modulus of these vectors
is $k_{min}=2\pi/L$. 

	The phonon angular frequencies derived by the HLR
method are obtained from the atomic fluctuations at the equilibrium
volume of the solid at a given temperature. This implies that the
volume dependence of the vibrational modes, i.e. the anharmonic effect
described by a QHA approach, is automatically included in the HLR
method. In addition, the HLR approach is able to describe anharmonic
effects beyond those described by a QHA. For example, 
in Ref. \onlinecite{ramirez05}
the HLR phonon frequencies of solid neon were compared with the QHA
result as a function of temperature in a range between 5-25 K, and
also with the harmonic approach. The error, with respect to the experimental
data, of the harmonic phonon frequency of the longitudinal phonon
at the point X of the Brillouin zone (BZ) of the fcc lattice amounts
to 40\%. The QHA phonon frequency was derived by diagonalization of
the dynamic matrix at the equilibrium volume of the solid at each
temperature. The QHA phonon frequency improves the harmonic result,
but is still about 20\% below the experimental data in the studied
temperature range. However, the HLR result is off by less than 2\%.
Then, the correlation between atomic fluctuations, as analyzed by
the HLR approach, is sensitive to anharmonic effects that are absent
in a QHA. 

\subsection{Phonon dispersion at 300 K}

\begin{figure}
\includegraphics[width=6.0cm]{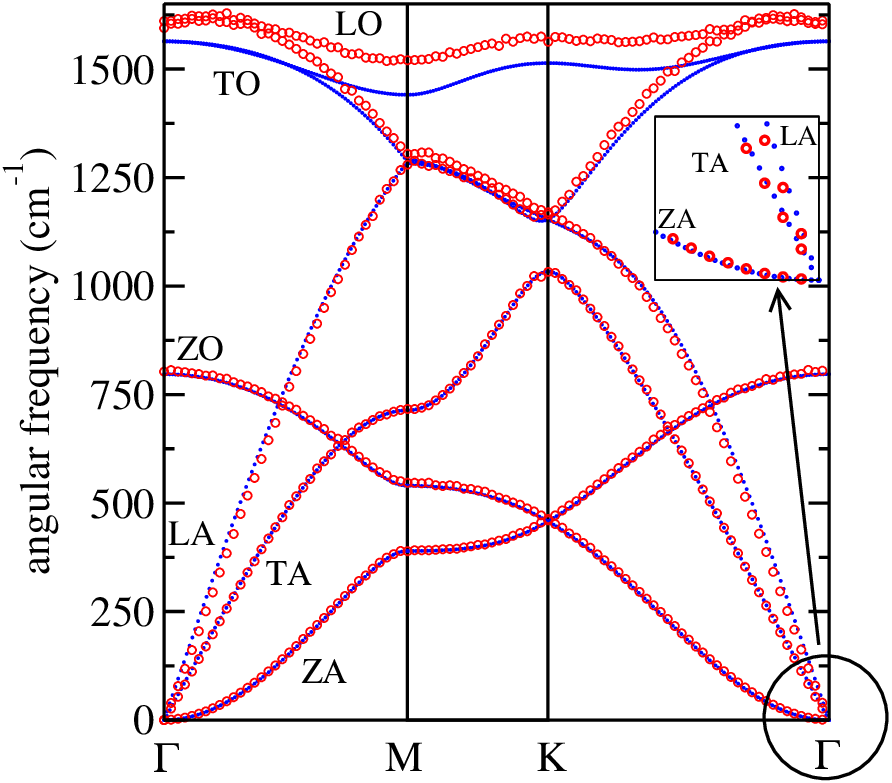}
\vspace{0.4cm}
\caption{The phonon dispersion relations of graphene at 300 K as derived by
the HLR method with the LCBOPII model are displayed as open circles.
The labels of the different branches are given. The inset shows the
acoustic region around the point $\varGamma$, that is studied in
this work. The harmonic limit of the LCBOPII model, as derived from
the diagonalization of the dynamical matrix, is displayed by closed
circles.}
\label{fig:w_k}
\end{figure}

The phonon dispersion relations of graphene have been derived by the
HLR method at 300 K and in-plane stresses, $\tau,$ in the range $[-0.02,0.01]$
eV/$\textrm{Å}^{2}$. For $\tau=0$, the frequencies, $\omega_{j}(\mathbf{k})$,
along the symmetry directions in the hexagonal Brillouin zone (BZ)
are displayed as open circles in Fig. \ref{fig:w_k}. The density
of points in the $\mathbf{k}-$grid is determined by the size of the
simulation cell and the shortest distance between $\mathbf{k}-$points
amounts to $k_{min}=0.042\;\textrm{Å}^{-1}$. The average in-plane
area at 300 K is $A_{p}=2.6162\;\textrm{Å}^{2}$/atom. The harmonic
phonon dispersion relations, as derived from the diagonalization of
the dynamical matrix using a denser grid of $\mathbf{k}-$points,
are displayed by closed circles. In the figure, these points appear
as a continuous line. The in-plane area corresponding to the minimum
potential energy amounts to $A_{p}=2.6189\;\textrm{Å}^{2}$/atom.
The phonon dispersion curves consist of three acoustic (A) and three
optical (O) bands. The atomic displacement vectors of the phonons
are either in-plane longitudinal (L), in-plane transverse (T), or
out-of-plane (Z). The linear dispersion of the LA and TA modes in
the long-wavelength limit ($k\rightarrow0)$, typical for the acoustic
modes of 3D solids, contrasts with the $k^{2}-$dependence of the
ZA mode in the harmonic approximation.\citep{zimmermann_08} Differences
between harmonic and HLR dispersion curves are due to anharmonic effects.
We are particularly interested in the anharmonicity of the long wavelength
acoustic region (LA, TA, and ZA branches), close to the special point
$\varGamma,$ a region shown by the inset of Fig. \ref{fig:w_k}.
The largest anharmonic effect in the dispersion relations at 300 K
corresponds to a blueshift of the optical branches (LO and TO). This
anharmonic effect has been studied in Ref. \onlinecite{ramirez_19}
with the conclusion that it is an artifact of the employed LCBOPII
model.

\section{Anharmonicity in the QHA \label{sec:QHA-approach}}

	In this Section, predictions of the QHA are presented
for graphene, as derived with the employed LCBOPII model. The QHA
analysis here focuses on the determination of the signs ($+$ or $-$)
expected for the anharmonic shifts of the acoustic LA, TA, and ZA
vibrational modes as a function of the applied in-plane stress and
temperature. In Sec. \ref{sec:Explicit-anharmonicty}, the QHA expectation
will be compared to the actual anharmonicity as derived by the HLR
method. As a result of this comparison, it will be clear that the
QHA does not provide a realistic description of the anharmonic shifts
found for these modes. For this reason, the analysis of the QHA here
is mainly done in qualitative terms.

The dependence of the vibrational frequencies, $\omega,$ with the
in-plane stress, $\tau$, and the temperature, $T$, is described
within the QHA by the Gruneisen constants. The Gruneisen constant
for a mode $w$ is defined as
\begin{equation}
\gamma_{\omega}=-\frac{A_{p}}{\omega}\frac{\partial\omega}{\partial A_{p}}\:.\label{eq:gamma_w}
\end{equation}
By considering the definition of the in-plane compressional modulus
(the 2D analogous to the bulk modulus of 3D solids),
\begin{equation}
B=-A_{p}\frac{\partial\tau}{\partial A_{p}}\:,\label{eq:B}
\end{equation}
the change of $w$ with the in-plane stress, $\tau,$ is expressed
as a function of the Gruneisen constant as,
\begin{equation}
\frac{\partial\omega}{\partial\tau}=\frac{\partial\omega}{\partial A_{p}}\frac{\partial A_{p}}{\partial\tau}=\frac{\gamma_{\omega}\omega}{B}\:.\label{eq:deri_w_tau}
\end{equation}
Analogously, the change of $w$ with $T$ can be deduced by considering
the definition of $\gamma_{w}$ and the thermal expansion coefficient,
$\alpha_{T}=A_{p}^{-1}\partial A_{p}/\partial T$, as
\begin{equation}
\frac{\partial\omega}{\partial T}=\frac{\partial\omega}{\partial A_{p}}\frac{\partial A_{p}}{\partial T}=-\gamma_{\omega}\omega\alpha_{T}\:.\label{eq:deri_w_T}
\end{equation}

\begin{figure}
\includegraphics[width=6.0cm]{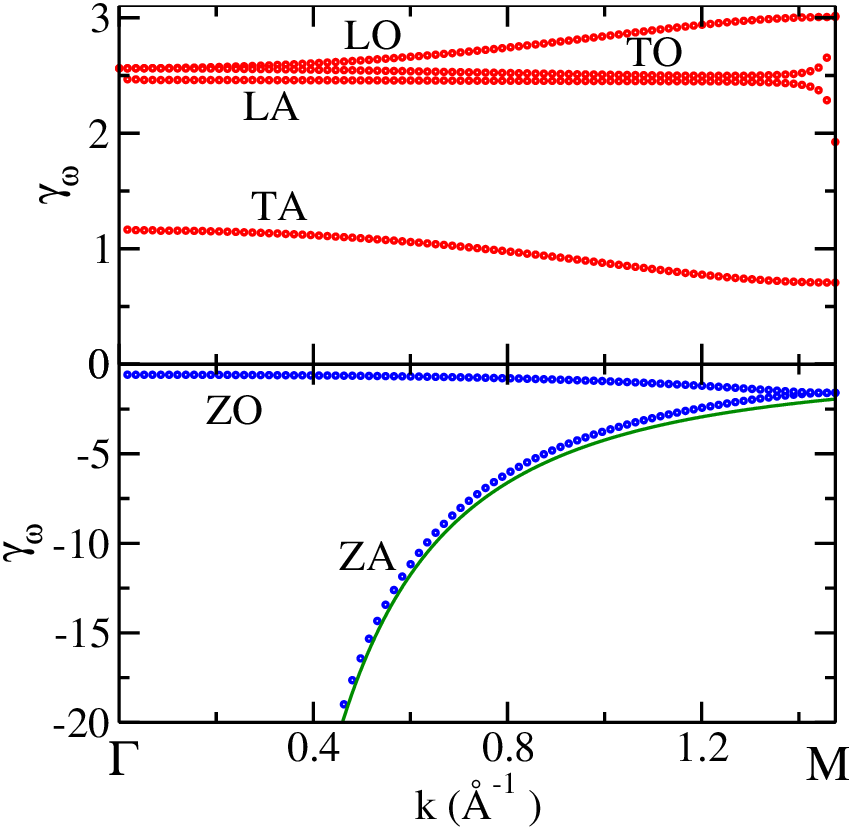}
\vspace{0.4cm}
\caption{The Gruneisen constants of the six phonon dispersion bands of graphene
along the $\varGamma M$ symmetry direction are shown by closed circles.
The bands with $z-$polarizations (ZO, ZA) display $\gamma_{\omega}<0$.
The continuous line is the long-wavelength limit of $\gamma_{ZA}$,
as given by Eq. (\ref{eq:gamma_ZA}) for $\tau=0$. The data correspond
to the LCBOPII model. Note the different vertical scales in the upper
and lower parts of the figure.}
\label{fig:gamma_k}
\end{figure}

The Gruneisen constants of graphene along the symmetry direction $\varGamma M$
are presented in Fig. \ref{fig:gamma_k}. $\gamma_{\omega}$ was calculated
with Eq. (\ref{eq:gamma_w}) by a numerical differentiation of the
harmonic frequencies of the LCBOPII model using a relative increase
in the equilibrium area, $A_{p}$, of 0.2\%. The LCBOPII results for
$\gamma_{\omega}$ are in good agreement with the ab-initio calculation
of Ref. \onlinecite{mounet05}. The vibrational bands LO, TO, LA,
and TA display $\gamma_{\omega}>0$, i.e., the usual behavior found
in most solids. These bands exhibit $(x,y)-$polarization. The two
$z-$bands (ZO, ZA) display however, $\gamma_{\omega}<0$. 

The Gruneisen constants of the ZA modes diverge in the limit $k\rightarrow0$.
The QHA dispersion relation of the ZA band in the long-wavelength
limit is given as\citep{pedro12a}
\begin{equation}
\rho\omega_{ZA}^{2}=-\tau k^{2}+\kappa k^{4}.\label{eq:rho_w2_ZA_QHA}
\end{equation}
The $k^{2}-$term vanishes for an unstressed layer ($\tau=0$) at
$T\rightarrow0$. By taking the $\tau-$derivative of the last expression
and with the help of Eq. (\ref{eq:deri_w_tau}), one gets 
\begin{equation}
\gamma_{ZA}=\frac{B}{2\tau-2\kappa k^{2}}\:.\label{eq:gamma_ZA}
\end{equation}
The Gruneisen constant $\gamma_{ZA}$ displays a $k^{-2}$ divergence
when $\tau=0$, in agreement with Ref. \onlinecite{pedro12a}. The
long-wavelength approximation for $\gamma_{ZA}$ in Eq. (\ref{eq:gamma_ZA})
is plotted by a continuous line in Fig. \ref{fig:gamma_k}, where
we have used the bending constant ($\kappa=1.5$ eV) and the in-plane
stiffness ($B=12.6$ eV/$\textrm{Å}^{2}$) corresponding to the harmonic
limit at $\tau=0$. The approximation for $\gamma_{ZA}$ is rather
realistic in the whole BZ.

\begin{table}
\caption{Signs of the expected QHA shifts of the vibrational frequencies of
the acoustic LA, TA, and ZA modes in graphene. $+$($-$) indicates
a blueshift (redshift) of the frequency of the modes. The last two
columns give the actual anharmonic shifts as derived from the HLR
approach in Sec. \ref{sec:Explicit-anharmonicty}. }
\vspace{0.5cm}
\begin{tabular}{|c||c|c|c||c|c|}
\cline{2-6}
\multicolumn{1}{c||}{} & \multicolumn{3}{c||}{QHA} & \multicolumn{2}{c|}{HLR}\tabularnewline
\cline{2-6}
\multicolumn{1}{c||}{} & $\gamma_{\omega}$ & $\frac{\partial\omega}{\partial T}$ & $\frac{\partial\omega}{\partial\tau}$ & $\frac{\partial\omega}{\partial T}$ & $\frac{\partial\omega}{\partial\tau}$\tabularnewline
\hline
LA/TA & $>0$ & $+$ & $+$ & $-$ & $-$\tabularnewline
\hline
ZA & $<0$ & $-$ & $-$ & $+$ & $-$\tabularnewline
\hline
\end{tabular}
\label{Tab:1}
\end{table}

The QHA predictions for the sign ($+$ or $-$) of the frequency shifts
of the acoustic branches of graphene with both temperature, $T$,
and in-plane stress, $\tau$, are summarized in Tab. \ref{Tab:1}.
The results are derived from Eqs. (\ref{eq:deri_w_tau}) and 
(\ref{eq:deri_w_T}).
The signs of the temperature shifts presented in
Tab. \ref{Tab:1} correspond to the case where the thermal expansion
coefficient in Eq. (\ref{eq:deri_w_T}) is $\alpha_{T}<0$, which
is the expected QHA behavior when $\gamma_{ZA}<0$.\citep{mounet05}

\section{Explicit anharmonicity \label{sec:Explicit-anharmonicty}}

In this Section, the anharmonic shifts of the acoustic vibrational
modes of graphene are analyzed by the HLR method. This method is sensitive
to both the volume dependent quasi-harmonic effect and the temperature
dependent explicit anharmonicity. The preponderance of explicit anharmonicity
will be identified by comparing the HLR results to the expectations
of the QHA. Firstly, the anharmonicity of the mode with largest wavelength
of each band, $\omega_{j}(\mathbf{k}_{min})$, ($j=$LA, TA, and ZA)
is studied. Secondly, the elastic coefficients of the layer are derived
from the $k-$dependence of the acoustic dispersion bands. 

\subsection{Acoustic modes with longest wavelengths\label{subsec:Acoustic-modes-with}}

\subsubsection{LA and TA modes}

\begin{figure}
\includegraphics[width=6.0cm]{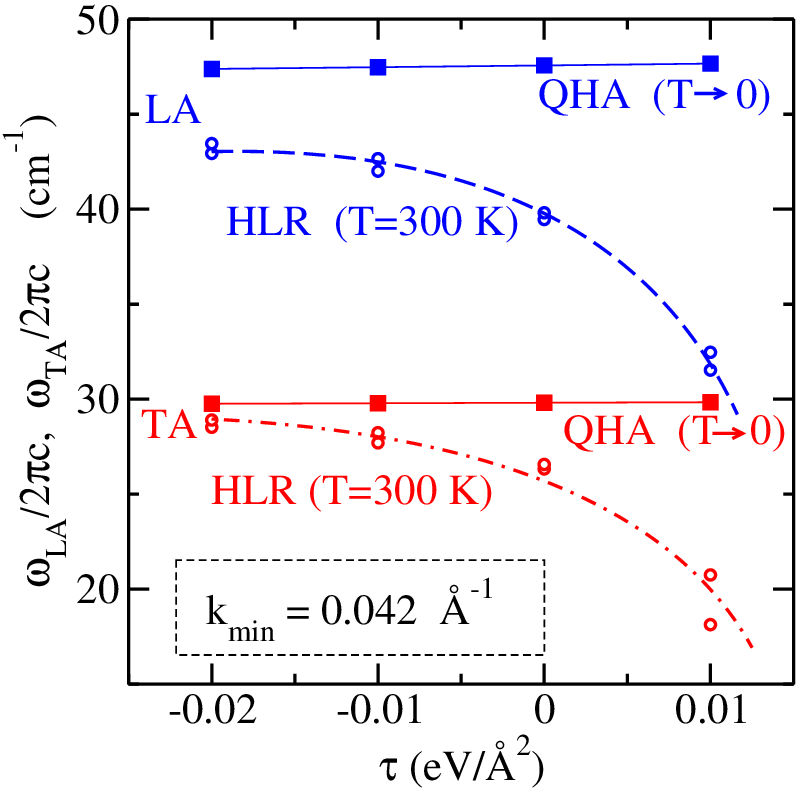}
\vspace{0.7cm}
\caption{Dependence of the angular frequency of the LA and TA phonons of 
graphene with the in-plane stress, $\tau$, for the two $\mathbf{k}-$points
with modulus $k_{min}$. The HLR wavenumbers at 300 K are given by
open circles. The QHA results in the $T\rightarrow0$ limit are shown
as closed squares. Lines are guides to the eye. }
\label{fig: w_k_min_LA_TA}
\end{figure}

For the employed simulation cell size, $k_{min}=2\pi/L=0.042\;\textrm{Å}^{-1}$.
There are two wavevectors with modulus $k_{min}$, oriented along
the $x-$ and $y-$axes.\citep{ramirez_19} The angular frequency
$\omega_{j}(\mathbf{k}_{min})$ for the in-plane polarized bands ($j=$
LA and TA) are displayed in Fig. \ref{fig: w_k_min_LA_TA} as a function
of the in-plane stress. The open circles show the HLR results at 300
K. The results for the $\mathbf{k}-$points $(k_{min},0)$ and $(0,k_{min})$
should be nearly identical, apart from the statistical error of the
simulation, as the layer appears isotropic in the long-wavelength
limit. At 300 K, one observes a sharp redshift of the LA/TA frequencies
as the layer is compressed ($\tau$ increases), that is indicated
with a negative sign in the HLR column ($\partial\omega/\partial\tau$)
of Tab. \ref{Tab:1}.

In a classical $T\rightarrow0$ limit, the temperature dependent explicit
anharmonicity vanishes, and the QHA becomes exact. This is only true
in a classical limit. In the real world, there appears explicit anharmonicity
even at $T\rightarrow0$, as consequence of the zero-point vibration.
In Fig. \ref{fig: w_k_min_LA_TA}, we have plotted the QHA frequencies
at $T\rightarrow0$ (closed squares). These values were derived by
numerical diagonalization of the dynamical matrix with the equilibrium
area at $T\rightarrow0$ for each $\tau$. The QHA angular frequencies
display a small blueshift as the in-plane stress $\tau$ increases
(see Tab. \ref{Tab:1}). This result contrasts with the redshift revealed
by the HLR method.

In the QHA, the frequency shift for the LA/TA modes is positive (blue-shift)
for rising temperature (see Tab. \ref{Tab:1}). However, one finds
that the HLR frequencies at 300 K are redshifted with respect to the
$T\rightarrow0$ limit. This behavior agrees with the anharmonicity
derived by perturbation theory for the in-plane LA/TA modes.\citep{amorim14}
Thus, the QHA alone is unable to predict correctly even the sign of
the frequency shifts for the LA/TA modes as a function of either the
temperature or the in-plane stress. This analysis points out the importance
of the explicit anharmonicity for the long-wavelength LA/TA modes
in graphene.

\subsubsection{ZA mode\label{subsec:ZA-mode}}

\begin{figure}
\includegraphics[width=6.0cm]{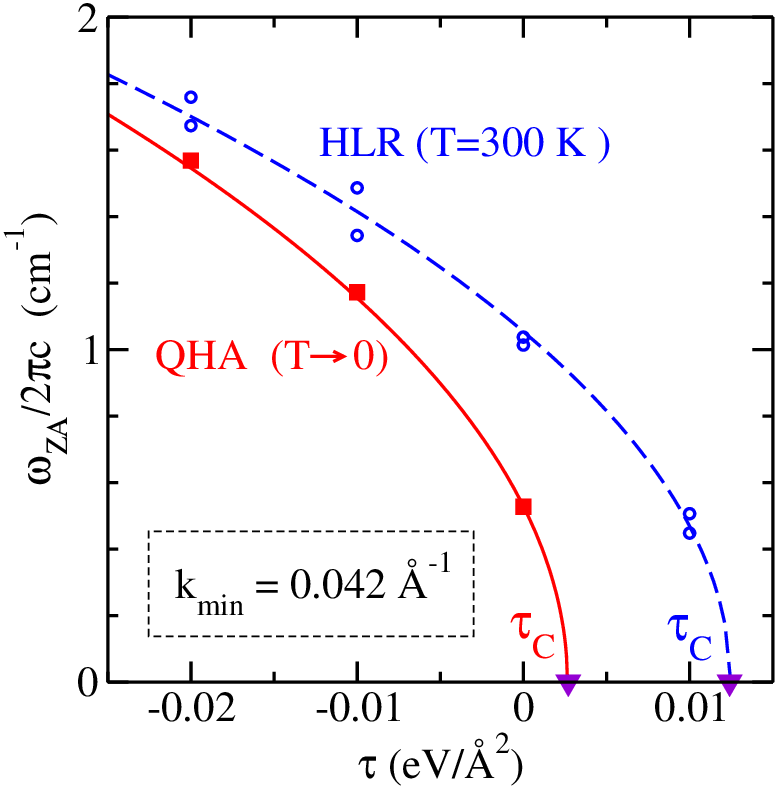}
\vspace{0.7cm}
\caption{Dependence of the angular frequency of the ZA phonons of graphene
with the in-plane stress, $\tau$, for the two $\mathbf{k}-$points
with modulus $k_{min}$. The HLR frequencies at 300 K are given by
circles. The broken line is a two-parameter fit ($d,\tau_{C}$) to
$\omega_{ZA}=d(\tau_{C}-\tau)^{1/2}$. The QHA results in the $T\rightarrow0$
limit, derived by diagonalization of the dynamical matrix, are shown
as squares. The continuous line corresponds to Eq. (\ref{eq:w_ZA_tau_C}).
The filled triangles represent the critical stress, $\tau_{c}$, where
the soft mode $\omega_{ZA}(k_{min})$ becomes unstable.}
\label{fig:w_k_min_ZA}
\end{figure}

In the low temperature limit ($T\rightarrow0$), the dispersion of
the ZA mode for long-wavelengths is given in the QHA by Eq. (\ref{eq:rho_w2_ZA_QHA}).
For the $\mathbf{k}-$vectors with modulus $k_{min}$, one gets,
\begin{equation}
\omega_{ZA}=k_{min}\left(\frac{\tau_{C}-\tau}{\rho}\right)^{1/2}\:.\label{eq:w_ZA_tau_C}
\end{equation}
For the cell with $N=8400$ atoms, the critical stress $\tau_{C}=\kappa k_{min}^{2}=3\times10^{-3}$
eV/$\textrm{Å}^{2}$. The critical stress $\tau_{C}$ signals the
point where the phonon mode becomes soft and its angular frequency
vanishes, $\omega_{ZA}(k_{min})=0$. The angular frequency $\omega_{ZA}(k_{min})$
as a function of $\tau$ given by Eq. (\ref{eq:w_ZA_tau_C}) is plotted
in Fig. \ref{fig:w_k_min_ZA} as a continuous line. The open squares
were calculated by diagonalization of the dynamical matrix for the
area $A_{p}$ in equilibrium at $T\rightarrow0$ for given stress
$\tau$. As $\tau$ increases up to the critical stress, $\tau_{C}$,
the mode $\omega_{ZA}(k_{min})$ becomes soft, and the finite flat
layer becomes mechanically unstable. The instability leads to a static
deformation via sinusoidal wrinkles with wavevector $\mathbf{k}_{min}$.\citep{ramirez18,ramirez_18b} 

At 300 K, the HLR angular frequencies $\omega_{ZA}(k_{min})$ are
displayed as a function of $\tau$ in Fig. \ref{fig:w_k_min_ZA} (open
circles). The ZA phonons are blueshifted with respect to the $T\rightarrow0$
limit. This contrasts with the redshift found for the LA/TA modes
by rising temperature in Fig. \ref{fig:w_k_min_ZA}. The blueshift
for the ZA modes implies that the critical stress, $\tau_{C}$, increases
with temperature. In other words, the mechanical stability of the
flat layer increases as temperature rises. The critical stress amounts
to $\tau_{C}=1.2\times10^{-2}$ eV/$\textrm{Å}^{2}$ at 300 K. It
is important to recall that the value of $\tau_{C}$ displays a significant
finite size effect, as $\tau_{C}$ depends on $k_{min}=2\pi/L$.
	The calculation of the critical stress for additional
cell sizes with $N=$ 1500 and 960 atoms, gives the result $\tau_{C}=3.2\times10^{-2}$
and $5.4\times10^{-2}$ eV/$\textrm{Å}^{2},$ respectively. These
values imply that the finite size effect in $\tau_{C}$ has a $N^{-1}$
dependence.  The smaller the size of the simulation
cell, the larger the stability of the flat morphology of the layer.\citep{ramirez17}

The shifts of the frequency of the ZA modes found by the HLR method
by rising either the temperature (blueshift) or the in-plane stress
(redshift) are summarized in Tab. \ref{Tab:1}. The QHA predicts effects
of opposite sign to the HLR method in the three acoustic bands of
graphene (LA, TA, and ZA) when the temperature increases. However,
the frequency shift of the ZA modes with the in-plane stress $\tau$
has, in the QHA, the same sign as in the HLR method.

\subsection{LA, TA dispersions}

\begin{figure}
\includegraphics[width=6.0cm]{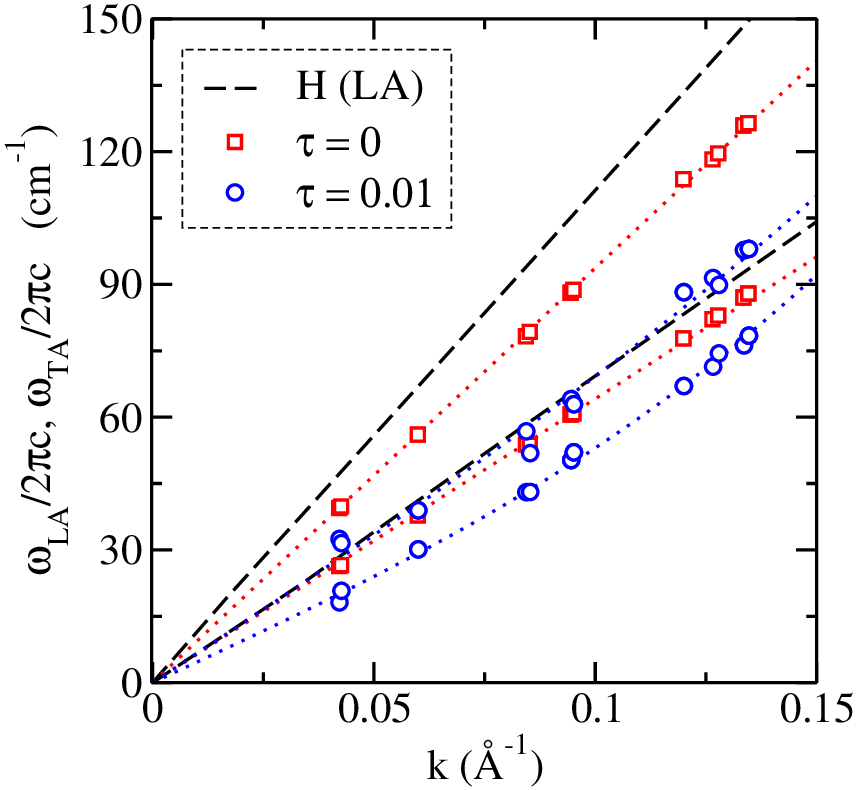}
\vspace{0.7cm}
\caption{LA and TA dispersion relations of graphene at in-plane stress $\tau=0$
(open squares) and 0.01 eV/$\textrm{Å}^{2}$ (open circles) . The
two bands can be distinguished as the slope of the LA branch is larger.
The dotted lines are least square fits of the simulation data to the
analytic function in Eq. (\ref{eq:LA_TA_fit}). The harmonic result
at $\tau=0$ is displayed by broken lines. At the compressive stress
$\tau=0.01$ eV/$\textrm{Å}^{2}$ the layer is close to its limit
of mechanical stability. }
\label{fig:LA_TA_disp}
\end{figure}

\begin{figure}
\includegraphics[width=6.0cm]{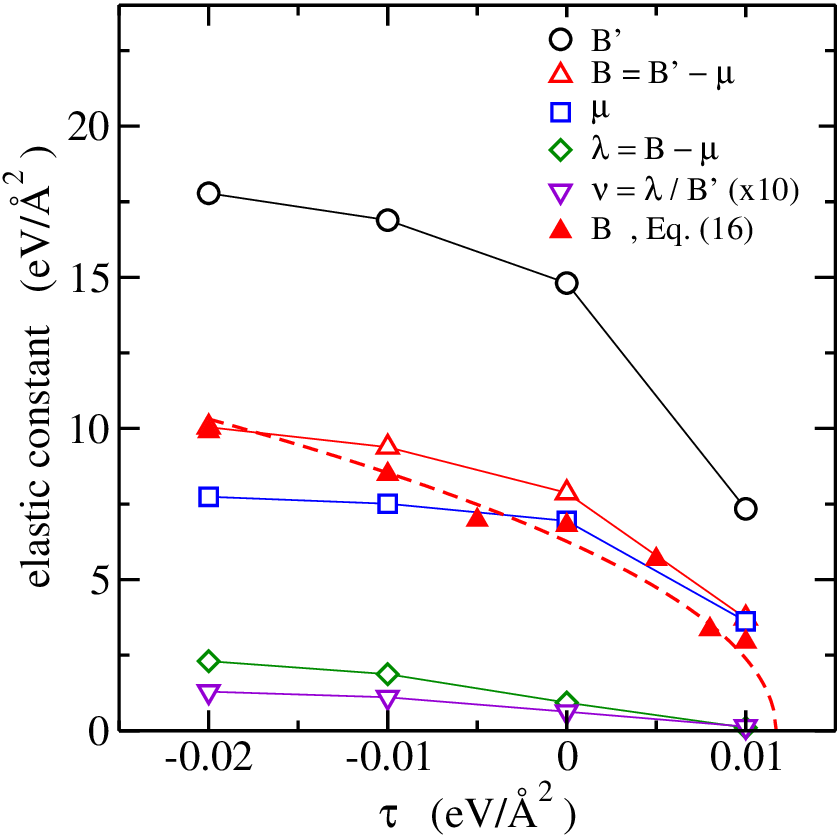}
\vspace{0.4cm}
\caption{In-plane stress dependence of the elastic constants of graphene at
300 K as derived from the slopes of the LA and TA phonon dispersion
bands in the long-wavelength limit. Shown are the unilateral compressional
modulus ($B'$), the shear modulus ($\mu$), the in-plane compressional
modulus ($B$), the Lamé's first coefficient ($\lambda$), and the
Poisson's ratio ($\nu$). The in-plane compressional modulus, derived
from the fluctuation formula of the $N\tau T$ ensemble {[}Eq. (\ref{eq:B_fluc}){]},
is displayed by closed triangles. The broken line is a least squares
fit of the closed triangles by Eq. (\ref{eq: B_tau_cri_fit}). The
continuous lines are guides to the eye.}
\label{fig:in_plane_elas_cte}
\end{figure}

The acoustic LA and TA phonon dispersions at two in-plane stresses
($\tau=0$ and 0.01 eV/$\textrm{Å}^{2}$) are displayed for $\mathbf{k}-$vectors
with $k<0.15$ eV/$\textrm{Å}^{2}$ in Fig. \ref{fig:LA_TA_disp}.
For comparison, the LA/TA harmonic phonon dispersions at $\tau=0$
are also plotted. The anharmonicity caused by increasing the temperature
and the in-plane stress is reflected by red-shifts of the LA/TA dispersion
curves with respect to the harmonic ones. Longitudinal ($v_{L}$)
and transverse ($v_{T}$) sound velocities were derived from the slope
of the dispersion curves in the long-wavelength limit ($k\rightarrow0$)
by a least squares fit of the function
\begin{equation}
\omega=\left(v^{2}k{{}^2}+f\:k^{3}\right)^{1/2}\:,\label{eq:LA_TA_fit}
\end{equation}
where $v$ is the sound velocity, and $f$ is a fitting constant.
For an isotropic elastic layer, the sound velocities are related to
the elastic constants by
\begin{equation}
v_{T}=\left(\frac{\mu}{\rho}\right)^{1/2}\:,
\end{equation}
\begin{equation}
v_{L}=\left(\frac{B'}{\rho}\right)^{1/2}\:,
\end{equation}
where $\mu$ is the shear modulus (Lamé's second coefficient), and
$B'$ is the unilateral compressional modulus defined as,\citep{behroozi_96}
\begin{equation}
B'=\lambda+2\mu\:,
\end{equation}
with $\lambda$ being the Lamé's first coefficient. The in-plane compressional
modulus is derived from $\mu$ and $B'$ as
\begin{equation}
B=B'-\mu=\lambda+\mu\:,\label{eq:B_disp}
\end{equation}
while the Poisson's ratio can be obtained from $\nu=\lambda/B\text{' }$.\citep{behroozi_96}
The in-plane stress dependence of the elastic moduli, $\mu,$ $B'$,
$B$, $\lambda,$ and $\nu$ of graphene, as derived from the HLR
analysis is presented in Fig. \ref{fig:in_plane_elas_cte}. The Poisson's
ratio has been scaled by a factor of 10 to be visible in the plot.
All elastic constants display a sharp decrease with positive (compressive)
in-plane stress. The Lamé's first coefficient $\lambda$ takes the
value 0.9 eV/$\textrm{Å}^{2}$ at $\tau=0$, while it is reduced to
$\lambda=0.1$ eV/$\textrm{Å}^{2}$ at $\tau=10^{-2}$ eV/$\textrm{Å}^{2}$.
This compressive stress is close to the limit of mechanical stability
of the employed simulation cell at 300 K, $\tau_{C}=1.2\times10^{-2}$
eV/$\textrm{Å}^{2}$. The Poisson's ratio $\nu$, proportional to
the Lamé's first coefficient $\lambda$, remains positive in the whole
range of studied in-plane stresses. In Tab. \ref{Tab:2} we summarized
the results for the Lamé's coefficients of graphene at 300 K.

\begin{table}
\caption{The in-plane area, $A_{p}$, the elastic moduli, $\mu$ and $\lambda$,
the surface tension, $\sigma$, and the bending rigidity, $\kappa$,
of graphene as derived from $N\tau T$ simulations at $T=300$ K and
$N=8400$ atoms. The harmonic limit is also given. }
\vspace{0.5cm}
\begin{tabular}{|c|c|c|c|c|c|c|}
\cline{2-7}
\multicolumn{1}{c|}{} & $\tau$ (eV/$\textrm{Å}^{2}$) & 
    $A_{p}$($\textrm{Å}^{2}$/atom) & $\mu$(eV/$\textrm{Å}^{2}$) & 
    $\lambda$(eV/$\textrm{Å}^{2}$) & $\sigma$ (eV/$\textrm{Å}^{2}$) & 
    $\kappa$ (eV)\tabularnewline
\hline
H & 0 & 2.6189 & 9.3 & 3.3 & 0 & 1.49\tabularnewline
\hline
\hline
$N\tau T$ & 0.01 & 2.6103 & 3.6 & 0.1 & -0.006 & 1.70\tabularnewline
\hline
$N\tau T$ & 0 & 2.6162 & 6.9 & 0.9 & 0.008 & 1.70\tabularnewline
\hline
$N\tau T$ & -0.01 & 2.6195 & 7.5 & 1.8 & 0.017 & 1.69\tabularnewline
\hline
$N\tau T$ & -0.02 & 2.6223 & 7.7 & 2.3 & 0.025 & 1.71\tabularnewline
\hline
\end{tabular}
\label{Tab:2}
\end{table}

	Experimental phonon dispersion relations of a quasi-freestanding
graphene sample grown on Pt(111), characterized by the weakest graphene-metal
interaction, have been determined by high-resolution electron energy
loss spectroscopy (HREELS). The sound velocities derived from the
slopes of the TA and LA branches were 14.0 and 22.0 km/s, 
respectively.\citep{politano12}
An optical technique to probe the acoustic TA and LA phonon branches
of graphene near the $\varGamma$ point by double resonant Raman scattering
reported acoustic sound velocities of 12.9 and 19.9 km/s, 
respectively.\citep{cong_19}
The sound velocities in Fig. \ref{fig:in_plane_elas_cte} for the
unstressed graphene layer ($\tau=0)$ are 12.1 and 17.7 km/s, respectively,
in reasonable agreement to the previous experimental data.

It is interesting to compare the compressional modulus, $B$ (open
triangles in Fig. \ref{fig:in_plane_elas_cte}) derived from the LA/TA
phonon dispersion relations, to the value derived from the fluctuation
formula corresponding to the $N\tau T$ ensemble,\citep{landau_80,ramirez17} 
\begin{equation}
B=\frac{A_{p}k_BT}{N\delta A_{p}^{2}}\:.\label{eq:B_fluc}
\end{equation}
The quadratic deviation $\delta A_{p}^{2}$ is the ensemble average
$\left\langle A_{p}^{2}\right\rangle -\left\langle A_{p}\right\rangle ^{2}$.
The closed triangles in Fig. \ref{fig:in_plane_elas_cte} display
$B$ calculated by the fluctuation formula at several in-plane stresses
for MD simulations using cells with $N=8400$ atoms. These MD results
for the compressional modulus $B$ seem to be systematically smaller
than those derived via Eq. (\ref{eq:B_disp}). This difference was
already observed in Ref. \onlinecite{ramirez_19} for simulations
cells with $N=960$ atoms, and was attributed to the out-of-plane
fluctuations of the layer, that make the layer to loose a strict 2D
character. 

For a 2D solid approaching a mechanical instability, the compressional
modulus is expected to display the following dependence with the in-plane
stress, $\tau$, (see Appendix \ref{sec:Spinodal-relations})
\begin{equation}
B=b(\tau_{C}-\tau)^{1/2}\:,\label{eq: B_tau_cri_fit}
\end{equation}
where $b$ is a constant, and $\tau_{C}$ is the critical stress for
the instability to occur. The broken line in Fig. \ref{fig:in_plane_elas_cte}
displays a least squares fit of Eq. (\ref{eq: B_tau_cri_fit}) to
the simulation data (closed triangles). The fitted parameters amount
to $b=58$ eV$^{1/2}$/$\textrm{Å}$ and $\tau_{C}=1.2\times10^{-2}$
eV/$\textrm{Å}^{2}$. This new independent estimation of the critical
stress, $\tau_{C}$, is in agreement with the value derived in Fig.
\ref{fig:w_k_min_ZA} by studying the soft mode, $\omega_{ZA}(k_{min})$,
of the ZA band with the HLR method.

\subsection{ZA dispersion }

\begin{figure}
\includegraphics[width=6.0cm]{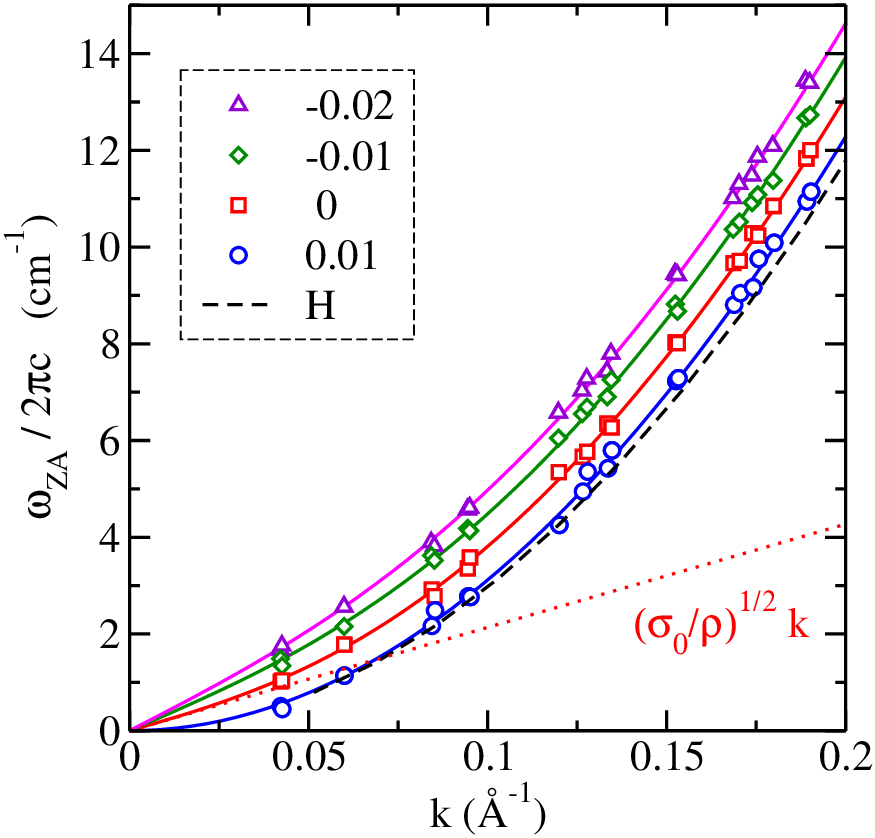}
\vspace{0.4cm}
\caption{ZA dispersion relations of graphene at 300 K as a function of the
in-plane stress ($\tau$) as derived with the LCBOPII model by the
HLR method. The values of $\tau$ are in eV/$\textrm{Å}^{2}$. The
continuous lines are the least squares fit of the simulation data
to the analytic function in Eq. (\ref{eq:ZA_fit}). The dotted line
shows the linear term of the fitted function for $\tau=0$ (open squares).
The broken line is the harmonic limit of the LCBOPII model derived
by diagonalization of the dynamical matrix at $\tau=0$. At the compressive
stress $\tau=0.01$ eV/$\textrm{Å}^{2}$ the flat layer is close to
its limit of mechanical stability. }
\label{fig:ZA_disp}
\end{figure}

The ZA phonon band of graphene for wavevectors with $k<0.2$ $\textrm{Å}^{-1}$
is displayed as a function of the in-plane stress at a temperature
of 300 K in Fig. \ref{fig:ZA_disp}. In Subsec. \ref{subsec:Acoustic-modes-with},
the anharmonic shifts found in the frequency of the ZA mode were a
blue-shift by rising temperature and a red-shift as the in-plane stress
increases (see Tab. \ref{Tab:1}). The ZA dispersion band has been
fitted to the expression,
\begin{equation}
\rho\omega_{ZA}^{2}=\sigma k^{2}+\kappa k^{4}+dk^{6}\:,\label{eq:ZA_fit}
\end{equation}
where the surface tension $\sigma$, the bending rigidity $\kappa$,
and $d$ are fitting parameters. The least squares fit included all
wavevectors with $k<0.3$ $\textrm{Å}^{-1}$. In the QHA at $T\rightarrow0$,
the relation between the surface tension $\sigma$ and the in-plane
stress $\tau$ is $\sigma=-\tau$ {[}see Eq. (\ref{eq:rho_w2_ZA_QHA}){]}.
It is interesting to study to which extent the temperature dependent
explicit anharmonicity modifies this QHA relation. 

The results of the least squares fits are plotted by continuous lines
in Fig. \ref{fig:ZA_disp}. The fluctuation tension, $\sigma_{0}$,
for the unstressed layer ($\tau=0$) is finite and amounts 
to $\sigma_{0}=8\times10^{-3}$ eV/$\textrm{Å}^{2}$, 
that translates into a finite sound velocity
$v_{ZA}=(\sigma_{0}/\rho)^{1/2}=0.4$ km/s. The linear term $v_{ZA}k$
of the ZA dispersion band at $\tau=0$ is displayed by a dotted line
in Fig. \ref{fig:ZA_disp}. The harmonic limit at $\tau=0$, as derived
from the diagonalization of the dynamical matrix, is shown by a broken
line in Fig \ref{fig:ZA_disp}.

\begin{figure}
\hspace{-0.6cm}
\includegraphics[width=6.0cm]{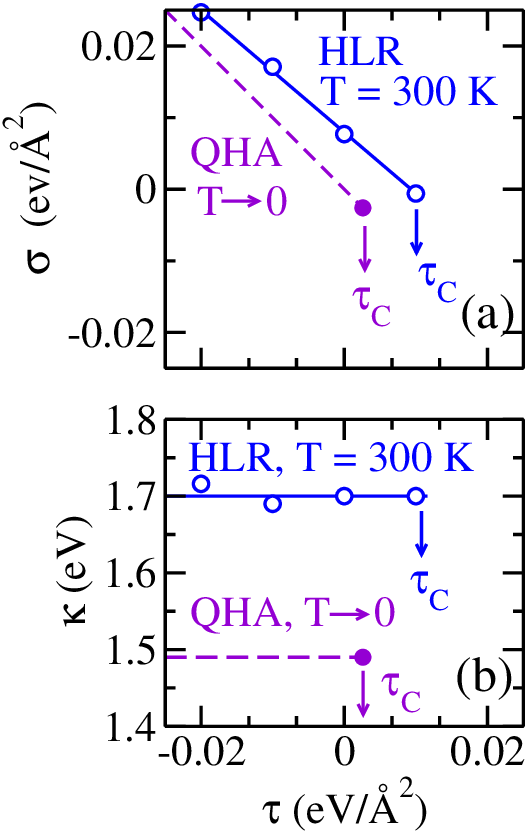}
\vspace{0.6cm}
\caption{$(a)$ The fluctuation tension, $\sigma,$ is displayed as a function
of the in-plane stress $\tau$. The broken line is the QHA at $T\rightarrow0$
K, $\sigma=-\tau$. The open circles are results from the ZA dispersion
relations derived by the HLR method at 300 K. The explicit anharmonicity
of graphene is the origin that the curve at 300 K is shifted with
respect to the $T\rightarrow0$ limit. The critical stress $\tau_{C}$
corresponding to the employed cell size ($N=8400)$ is indicated by
an arrow and a closed circle. $\tau_{C}$ is larger at 300 K than
in the $T\rightarrow0$ K limit. $(b)$ In-plane stress dependence
of the bending constant, $\kappa$, at 300 K and in the $T\rightarrow0$
limit. }
\label{fig: Z_elas_cte}
\end{figure}

The values of the fitted parameters, $\sigma$ and $\kappa$, are
summarized in Tab. \ref{Tab:2} and plotted as a function of the in-plane
stress in Fig. \ref{fig: Z_elas_cte}. At given stress $\tau$, the
fluctuation tension $\sigma$ is larger at 300 K than at $T\rightarrow0$.
The dependence of $\sigma$ with $\tau$ at 300 K is linear. A least
squares fit of the simulation results gives, 
\begin{equation}
\sigma=\sigma_{0}-0.9\tau\:.\label{eq:sigma}
\end{equation}
This relation is a consequence of the temperature dependent explicit
anharmonicity in graphene. The result contrast to the QHA relation
$\sigma=-\tau$ at $T\rightarrow0$. The bending rigidity $\kappa$
is at 300 K larger than at $T\rightarrow0$. $\kappa$ remains nearly
constant for the studied in-plane stresses, even at the compressive
stress of 0.01 eV/$\textrm{Å}^{2}$, close to the critical stress.
However, the other in-plane elastic moduli ($B'$ and $\mu)$ and
the fluctuation tension, $\sigma,$ display a significant variation
as a function of the in-plane stress $\tau$.

At 300 K, the critical stress, $\tau_{C}$, can be expressed with
the help of Eqs. (\ref{eq:ZA_fit}) and (\ref{eq:sigma}) as, 
\begin{equation}
\tau_{C}=\frac{\sigma_{0}+\kappa k_{min}^{2}}{0.9}\:.
\end{equation}
and takes the value $\tau_{C}=1.2\times10^{-2}$ for the fitted results
of $\sigma_{0}$ and $\kappa$. This is in agreement with the value
of $\tau_{C}$ derived at 300 K from the $\tau-$dependence 
of $\omega_{ZA}(k_{min}$) in Fig. \ref{fig:w_k_min_ZA}.

\begin{figure}
\includegraphics[width=6cm]{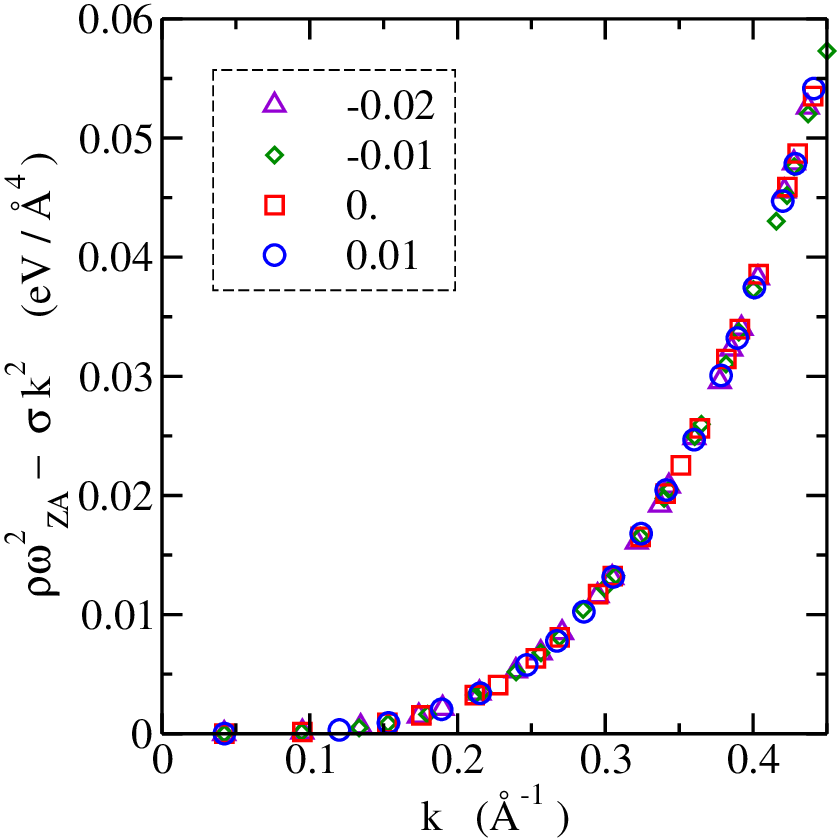}
\vspace{0.4cm}
\caption{The difference $\rho\omega_{ZA}^{2}-\sigma k^{2}$ is represented
as a function of $k$ for the four studied in-plane stresses. The
values of $\tau$ are given in eV/$\textrm{Å}^{2}$. The results for
the different stresses lie in the same curve. }
\label{fig:ZA_comp}
\end{figure}

For the range of in-plane stresses studied here, when $\tau$ changes,
the effect in the ZA band is limited to the $\sigma k^{2}-$term of
Eq. (\ref{eq:ZA_fit}). In Fig. \ref{fig:ZA_comp}, the difference,
$\rho\omega_{ZA}^{2}-\sigma k^{2}$, is displayed for the four simulated
stresses and wavevectors with $k<0.4$ $\textrm{Å}^{-1}$. The resulting
points lie all in the same curve, being nearly indistinguishable. 

\section{Comparison to analytical models\label{sec:Comparison-to-analytical}}

In this Section, the simulations results for the ZA dispersion are
compared with available analytical models. The prediction of the anomalous
exponent model in the long-wavelength limit is a renormalization of
the harmonic relation, $\rho\omega_{ZA}^{2}=\kappa k^{4}$, resulting
in the ZA phonon dispersion: $\rho\omega_{ZA}^{2}=\kappa_{A}k^{4-\eta}$,
with $\eta=0.82$ being an anomalous exponent.\citep{Los09,doussal17}
To quantify the agreement between this anharmonic model and the simulation
results, we performed a two-parameter least squares fit ($\kappa_{A},$$\alpha_{A}$)
of the points of the ZA bands shown in Fig. \ref{fig:ZA_disp} to
the function
\begin{equation}
\rho\omega_{ZA}^{2}=\kappa_{A}k^{\alpha_{A}}.\label{eq:anomalous_expo}
\end{equation}
The fitted region is the interval $[k_{min},k_{max}]$. While $k_{min}=0.042$
$\textrm{Å}^{-1}$ is a fixed value, determined by the size of the
simulation cell, $k_{max}$ is reduced from 0.6 to 0.1 $\textrm{Å}^{-1}$.
When $k_{max}=0.6$ $\textrm{Å}^{-1}$, the fit includes a large region
with 314 $\mathbf{k}-$points. When $k_{max}=0.1$ $\textrm{Å}^{-1}$,
the fit includes only the 10 $\mathbf{k}-$points with the longest
wavelengths. The change in the fitted parameters ($\kappa_{A},$$\alpha_{A}$)
as the value of $k_{max}$ decreases, allows us to visualize in which
way the simulation results converge to the long-wavelength limit given
by Eq. (\ref{eq:anomalous_expo}). 

\begin{figure}
\includegraphics[width=6.0cm]{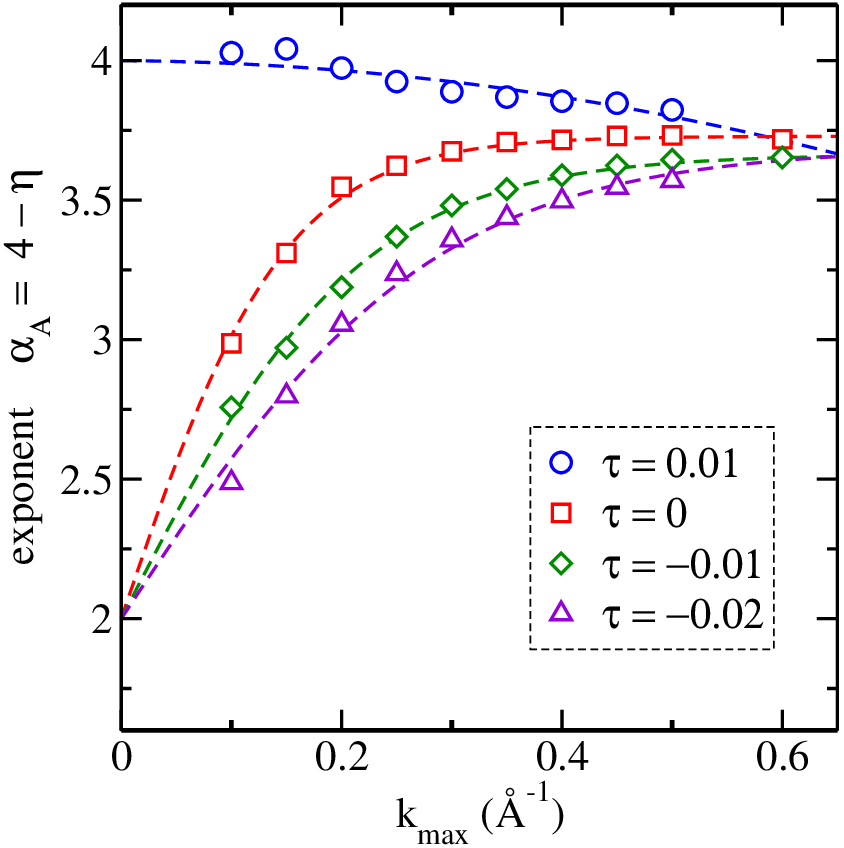}
\vspace{0.4cm}
\caption{Value of the exponent $\alpha_{A}$ derived from the fit 
of the simulation
results for $\rho\omega_{ZA}^{2}$ to the anomalous exponent model,
$\kappa_{A}k^{\alpha_{A}}$ {[}see Eq. (\ref{eq:anomalous_expo}){]}.
The numerical fit is performed including wavevectors with modulus
$k<k_{max}$. The results for the exponent  $\alpha_{A}$ are shown
as a function of $k_{max}$ to quantify the convergence of the simulation
results to the long-wavelength limit in Eq. (\ref{eq:anomalous_expo}).
$\tau$ values are given in eV/$\textrm{Å}{}^{2}$. The lines are
guides to the eye.}
\label{fig:test_anomalous}
\end{figure}

The result of the fitted exponent $\alpha_{A}=4-\eta$ as a function
of $k_{max}$ is presented in Fig. \ref{fig:test_anomalous}. When
$k_{max}$ is large the exponent $\alpha_{A}$ takes a value close
to 4 for the four studied in-plane stresses, implying that $\rho\omega_{ZA}^{2}$
displays an overall dispersion that looks like $k^{4}$ when $k<0.6$
$\textrm{Å}^{-1}$. As $k_{max}$ decreases, the exponent $\alpha_{A}$
for the in-plane stresses $\tau\leq0$ reduces monotonically towards
lower values, that seem to converge in the long-wavelength limit to
$\alpha_{A}=2,$ that would imply a long-wavelength dispersion for
$\rho\omega_{ZA}^{2}$ with a dominant $k^{2}-$dependence. We do
not see any qualitative difference between $\tau<0$ (tensile stress)
and $\tau=0$ (unstressed layer), only that the convergence towards
the limit $\alpha_{A}=2$ is faster when the in-plane stress becomes
more tensile. The data in Fig. \ref{fig:test_anomalous} do not provide
any evidence that the long-wavelength limit of the dispersion $\rho\omega_{ZA}^{2}$
when $\tau=0$ should behave as $k^{3.2}$, as predicted by the anomalous
exponent model. 

For the compressive in-plane stress, $\tau=0.01$ eV/$\textrm{Å}^{2}$,
the behavior of the exponent $\alpha_{A}$ is qualitatively different.
As $k_{max}$ becomes smaller the exponent $\alpha_{A}$ approaches
$4$. The interpretation of this behavior is that the compressed layer
is close to its limit of mechanical stability, where there appears
a soft phonon mode in the ZA band. This critical behavior is signalized
by a $\rho\omega_{ZA}^{2}$ dispersion with $k^{4}-$dependence. We
stress that here the $k^{4}-$dependence of $\rho\omega_{ZA}^{2}$
is a fingerprint of a soft phonon mode in the flat layer. It would
be misleading to associate this $k^{4}-$dependence close to the critical
stress, $\tau_{C},$ to an absence of anharmonic effects. In fact,
the blue-shift of the harmonic ZA modes at 300 K by the temperature
dependent explicit anharmonicity, is compensated by an anharmonic
red-shift when the in-plane stress becomes more compressive (see Tab.
\ref{Tab:1}). At the critical stress, $\tau_{C}$, both anharmonic
effects compensate each other, and the ZA dispersion takes the form
of a $k^{4}-$dependence, signalizing the appearance of a soft ZA
mode in the anharmonic layer.

\begin{figure}
\includegraphics[width=6.0cm]{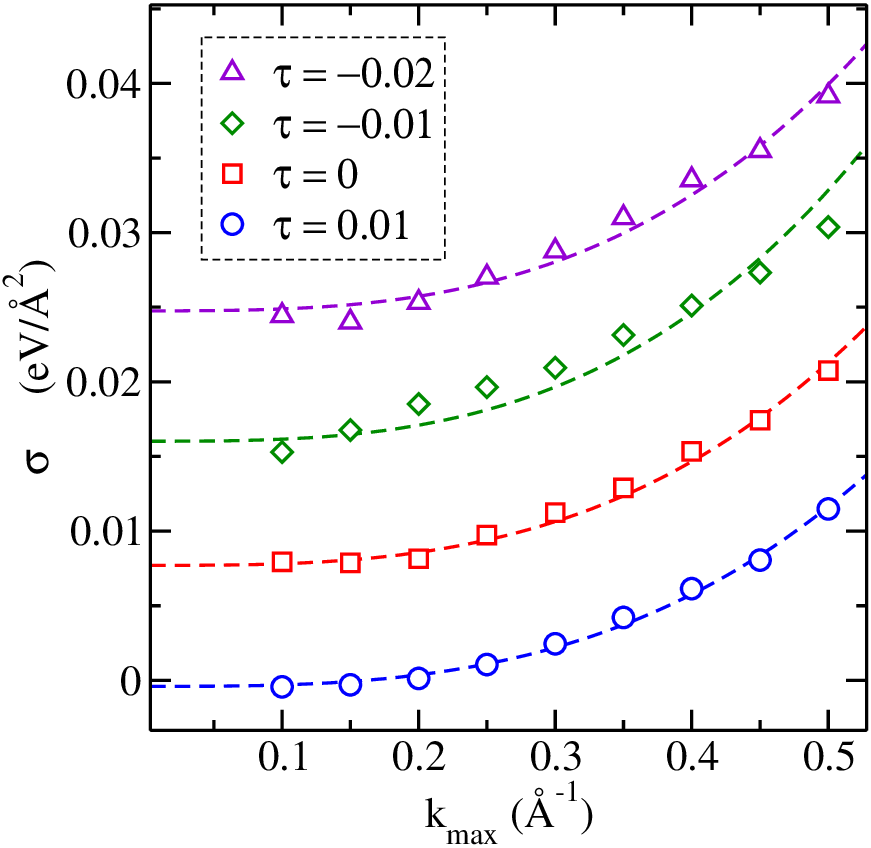}
\vspace{0.4cm}
\label{fig:check_pertur}
\caption{Value of the surface tension $\sigma$ from the fit 
of $\rho\omega_{ZA}^{2}$
to the prediction of the perturbation theory, $\sigma k^{2}+\kappa k^{4}$.
The numerical fit is performed using only ZA angular frequencies with
wavenumber modulus $k<k_{max}$. The results of $\sigma$ are shown
as a function of $k_{max}$ to study the convergence of the model
to the long-wavelength limit. $\tau$ values are given 
in eV/$\textrm{Å}{}^{2}$.
The lines are guides to the eye.}
\end{figure}

Perturbation theory predicts the long-wavelength limit of the ZA band
of graphene as $\rho\omega_{ZA}^{2}=\sigma k^{2}+\kappa k^{4}$. We
have tested this model against our simulation results by the same
method used to check the anomalous exponent model. Then, a two-parameter
least squares fit ($\sigma,\kappa$) of the ZA dispersion is made
in the interval $[k_{min},k_{max}]$. The result for the surface tension
$\sigma$ as a function of $k_{max}$ is displayed for the studied
in-plane stresses in Fig. 11. As $k_{max}$ decreases, $\sigma$ shows
a monotonic convergence towards a constant value in the long-wavelength
limit. At each studied in-plane stress, $\tau$, the relation $\sigma>-\tau$
is satisfied. This result differs from the QHA expectation, $\sigma=-\tau$,
because it is a consequence of explicit anharmonicity. 

Note that for the compressive stress, $\tau=0.01$ eV/$\textrm{Å}^{2}$,
$\sigma$ is negative. For finite size simulation cells, where $k_{min}=2\pi/L$
is finite, the ZA vibrational mode with lowest frequency $\rho\omega_{ZA}^{2}(k_{min})=\sigma k_{min}^{2}+\kappa k_{min}^{4}$
may be positive, even if $\sigma<0$ . It is at the critical stress,
$\tau_{C}$, that this mode becomes soft, $\rho\omega_{ZA}^{2}(k_{min})=0$
, and the flat surface morphology becomes mechanically unstable. In
the thermodynamic limit, $N\rightarrow\infty,$ the mechanical instability
would correspond to a vanishing value of the fluctuation  tension
($\sigma=0$), but for finite size cell the soft mode appears at negative
values of the surface tension, $\sigma<0$.

\section{Temperature and quantum effects\label{sec:Temperature-and-quantum}}

The analysis of the anharmonicity of the acoustic modes in graphene
at 300 K has been presented in the classical limit, using a relatively
large simulation cell, $N=8400$ atoms, and very long simulations
runs. With these conditions, quantum PI simulations at low temperature
would require an enormous computational effort. In a recent paper,
we have presented classical and quantum PI simulations of graphene
with a smaller simulation cell, $N=960$ atoms.\citep{ramirez18}
The quantum PIMD simulations were performed with in-plane stress $\tau=0$
and temperatures in the range $25-1000$ K. The smaller simulation
cell implies that the wavevector with smallest modulus is $k_{min}=0.12$
$\textrm{Å}^{-1}$, i.e., about three times larger than that one corresponding
to a cell with $N=8400$ atoms. The study of the long-wavelength limit
of the dispersion curves is less accurate with smaller cells, due
to the cut-off of all collective vibrations with wavelengths larger
than the cell dimension.

Nevertheless, the analysis of the long-wavelength limit of the ZA
band in Ref. \onlinecite{ramirez18} using a smaller cell is in good
in agreement with the analysis made in this paper. The fluctuation
tension $\sigma_{0}$ at 300 K was of $8.7\pm0.8$ meV/$\textrm{Å}^{2}$
with $N=960$ atoms, similar to our present result with a larger simulation
cell ($\sigma_{0}=$8 meV/$\textrm{Å}^{2}$). In the classical limit
$\sigma_{0}$ increases monotonously with temperature. The increase
is a consequence of the explicit anharmonicity and vanishes at $T\rightarrow0$
in a classical limit. The main quantum effect in the value of the
fluctuation tension $\sigma_{0}$ is that the anharmonicity is finite
even at $T\rightarrow0$, as a consequence of the zero-point vibration.
The extrapolated value at $T\rightarrow0$ is $\sigma_{0}\sim2.5$
meV/$\textrm{Å}^{2}$ in the quantum case.\citep{ramirez18} 

\begin{figure}
\includegraphics[width=6.0cm]{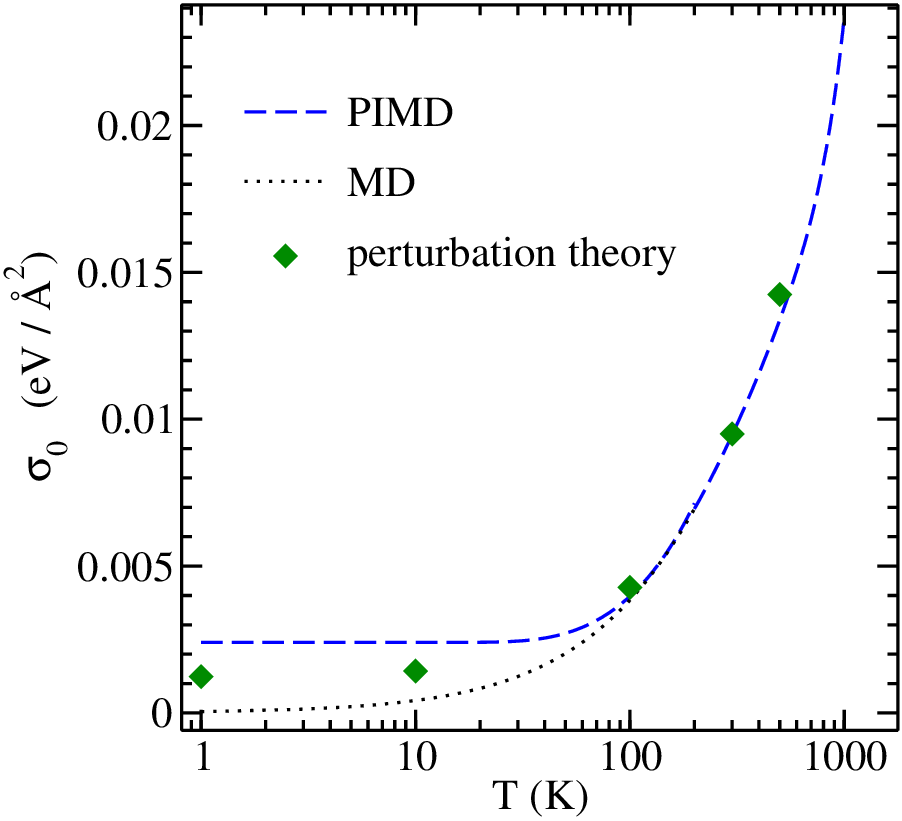}
\vspace{0.4cm}
\label{fig:quantum_effect}
\caption{Temperature dependence of the fluctuation tension, $\sigma_{0},$
of graphene at in-plane stress $\tau=0$. The broken line is a fit
of the quantum PIMD results with a simulation cell with $N=960$ atoms.
The dotted line represents the corresponding results in the classical
limit. Both results were taken from Ref. \onlinecite{ramirez18}.
The filled diamonds are the values of $\sigma_{0}$ derived from perturbation
theory in the quantum limit in Ref. \onlinecite{michel15}.}
\end{figure}

Further evidence of the agreement between our simulation results and
perturbation theory is presented in Fig. 12. We have plotted the fit
of the classical and quantum simulation results for $\sigma_{0}$
as a function of temperature, as derived in Ref. \onlinecite{ramirez18}
with a simulation cell with $N=960$ atoms. The temperature is presented
in logarithmic scale to highlight the difference between classical
and quantum results. This difference is significant only at temperatures
below 100 K. In addition, the results for the fluctuation tension
$\sigma_{0}$, that were derived in the quantum limit from perturbation
theory, are also plotted in Fig. 12.\citep{michel15} The absence
of any fitting parameter in this comparison between simulation and
perturbation theory results provides further evidence for their striking
agreement.

\section{Summary}

The anharmonicity of the acoustic phonon dispersion of graphene has
been studied in the long-wavelength limit with the HLR method. This
approach is based on the study of the correlation between the fluctuations
of atomic positions from their equilibrium values by means of computer
simulations. We have studied the phonon dispersion relations of graphene
at 300 K and at various in-plane stresses, from a tensile stress of
-0.02 eV/$\textrm{Å}^{2}$to a compressive stress of 0.01 eV/$\textrm{Å}^{2}.$
The latter is close to the mechanical stability limit of the flat
layer, where the phonons of the ZA dispersion band at $\varGamma$
become soft and cause the morphology of the flat layer to change by
formation of static sinusoidal wrinkles. The simulations were performed
at 300 K in the classical limit, which is a reasonable approach for
the acoustic phonon vibrations in the long-wavelength limit. These
are the modes having the lowest vibrational energies in the solid.

The QHA analysis of the LA/TA vibrational bands shows that this approach
is unable to predict the anharmonicity found for these modes. The
frequency of the long-wavelength limit of the LA/TA bands is predicted
by the QHA to be blue-shifted as either the temperature $T$ or the
in-plane stress $\tau$ increases. However the HLR approach shows
that the LA/TA frequencies are red-shifted by raising either $T$
or $\tau$. This is a consequence of the explicit anharmonicity of
these modes. The red-shift of the LA/TA vibrational bands with rising
temperature found in the simulation is in agreement with the prediction
of perturbation theory.\citep{amorim14}

The analysis of the long-wavelength limit of the ZA band reveals that
the QHA is unable to predict the blue-shift of the frequency of these
modes by increasing temperature. This blue-shift is an explicit anharmonic
effect with deep influence in the physical properties of the layer.
It is responsible for an increase in the mechanical stability of the
flat layer. Both zero-point atomic vibrations and a rise in temperature
produce a blue-shift of the ZA band in the long-wavelength limit.
However, a (compressive) increase of the in-plane stress $\tau$ of
the layer produces a red-shift of the frequency of the ZA modes. This
anharmonic effect has different sign than that one caused by temperature.
Therefore, the increased mechanical stability of the flat layer, caused
by an increase in $T$, can be compensated by the opposite effect
of rising the in-plane stress. An important conclusion is that for
an unstressed graphene membrane ($\tau=0)$ this compensation is not
perfect, and the layer displays a small but finite fluctuation tension
($\sigma_{0}>0)$ that determines the dispersion relation of the ZA
band in the long-wavelength limit as $\rho\omega_{ZA}^{2}=\sigma_{0}k^{2}$. 

The simulation results for the long-wavelength limit of the ZA band
have been compared with the predictions of two analytical models:
perturbation theory and the anomalous exponent model. Our results
are in good agreement with perturbation theory, that predicts a finite
fluctuation tension ($\sigma_{0}>0)$ in the ZA band and a red-shift
of angular frequencies in the LA/TA bands caused by finite temperature
and zero-point vibrations. The anomalous exponent model predicts a
vanishing sound velocity of the ZA phonons in the unstressed layer
that is not confirmed by our simulations.

\acknowledgments 

This work was supported by the Ministerio de Ciencia e Innovación
(Spain) through Grant PGC2018-096955-B-C44. We thank the support of
J. H. Los in the implementation of the LCBOPII model.

\appendix

\section{LCBOPII Potential\label{sec:LCBOPII-Potential}}

	The total binding energy, $\phi$, for a system consisting
of $N$ carbon atoms is defined with the LCBOPII empirical potential
as,\citep{los05}
\begin{eqnarray}
\phi & = & \frac{1}{2}  \sum_{i=1}^{N} \sum_{j=1}^{N} [
      S_{sr}^{down}(d_{ij})  V^{sr}(d_{ij})  \nonumber \\
  & + &  S_{lr}^{up}(d_{ij})  V^{lr}(d_{ij})  + 
    \frac{1}{Z_{i}^{mr}}  S_{mr}^{up}(d_{ij}) V^{mr}(d_{ij})
     ]  \: 
\end{eqnarray}
where $V^{sr}(d_{ij})$ describes short range and covalent interactions,
$V^{lr}(d_{ij})$ accounts for long range non-bonded interactions,
and $V^{mr}(d_{ij})$ represents the remainder of bonded (attractive)
interactions between atoms at middle range distances. $d_{ij}$ is
the interatomic distance between atoms $i$ and $j$. The prefactor
$1/Z_{i}^{mr}$ takes into account many-body effects, with $Z_{i}^{mr}$
being an effective middle range coordination number of atom $i$.
The switch functions $S_{sr}^{down},S_{lr}^{up}$, and $S_{mr}^{up}$
provide a smooth connection between the various interaction contributions.
The short range interaction, $V^{sr}(d_{ij})$, vanishes at distances
$d_{ij}>2.2$ $\textrm{Å}$ and describes both repulsive and atractive
pair potentials. The atractive term depends on bond order factors
that take into account many-body effects such as conjugation, presence
of anti-bonding states, and torsion. The long-range interaction $V^{lr}$
cuts off smoothly long-range interactions beyond 6 Å, while the middle
range attractive interactions $V^{lr}$ depend on bond angles and
on the presence of dangling bonds. For a detailed account of the analytical
structure and parameters of this empirical potential we refer to the
original work in Ref. \onlinecite{los05}. According to previous 
simulations,\citep{tisi_17}
the torsion parameters of the original LCBOPII model were slightly
modified to increase the bending constant of the graphene layer from
$\kappa=0.8$ to $1.5$ eV at $T\rightarrow0$. The last value displays
better agreement to experimental data and ab-initio calculations.\citep{lambin14}

	In our MD simulations of graphene we have derived
the potential energy $\phi$ of the layer with the LCBOPII model.
The atomic forces of the graphene configurations,
\begin{equation}
(F_{xi},F_{yi},F_{zi})=\left(-\frac{\partial\phi}{\partial x_{i}},-\frac{\partial\phi}{\partial y_{i}},-\frac{\partial\phi}{\partial z_{i}}\right),i=1,\ldots N,
\end{equation}
were calculated analitycally with the LCBOPII potential, as well as
the total derivative of the potential energy with respect the in-plane
area of the simulation cell, $d\phi/dA$. This derivative is defined
by the change of the potential energy of the simulation cell upon
an uniform isotropic strain in the $xy-$plane. These quantities are
required in the dynamic equations used to sample the $N\tau T$ ensemble.

\section{Dynamic Equations for the $N\tau T$ ensemble 
      \label{sec:Dynamic-Equations-for}}

The dynamic equations that generate the $N\tau T$
ensemble are reviewed here from the original literature in Refs. \onlinecite{martyna92,martyna94,ma96,tu98,ma99}.
In order to produce the isothermal-isobaric ensemble, the in-plane
area of the simulation cell, $A=NA_{p},$ is permitted to undergo
isotropic fluctuations. The employed extended system scheme treats
as dynamic variables the atomic positions $(x_{i},y_{i},z_{i})$ and
momenta $(p_{xi},p_{yi},p_{zi})$, with $i=1,\ldots,N$, the in-plane
area, $A$, and the momentum associated with the logarithm of the
in-plane area, $p_{A}$. In addition, chains of Nosé-Hoover thermostats
are employed to generate the thermal fluctuations of the distributed
positions $(x_{i},y_{i},z_{i})$, and momenta $(p_{xi},p_{yi},p_{zi})$.
An additional chain of thermostats is coupled to the ``barostat''
to control the area fluctuations. The equations of motion are\citep{tu98,ma99}

\begin{equation}
\dot{x}_{i}=\frac{p_{xi}}{m}+\frac{p_{A}}{W}x_{i}\;,
\end{equation}

\begin{equation}
\dot{z}_{i}=\frac{p_{zi}}{m}\;,
\end{equation}

\begin{equation}
\dot{p}_{xi}=F_{xi}-\left(1+\frac{1}{N}\right)\frac{p_{A}}{W}p_{xi}-\frac{p_{\xi xi1}}{Q}p_{xi}\;,
\end{equation}

\begin{equation}
\dot{p}_{zi}=F_{zi}-\frac{p_{\xi zi1}}{Q}p_{zi\;,}
\end{equation}

\begin{equation}
\dot{A}=\frac{2Ap_{A}}{W}\:,
\end{equation}

\begin{equation}
\dot{p}_{A}=2A(\tau_{int}-\tau)+\frac{1}{N}{\displaystyle \sum_{i=1}^{N}\left(\frac{p_{xi}^{2}+p_{yi}^{2}}{m}\right)}-\frac{p_{\beta1}}{Q_{\beta}}p_{A}\:,
\end{equation}

\begin{equation}
\dot{\xi}_{xij}=\frac{p_{\xi xij}}{Q}\:,
\end{equation}

\begin{equation}
\dot{p}_{\xi xi1}=\left(\frac{p_{xi}^{2}}{m}-k_{B}T\right)-p_{\xi xi1}\frac{p_{\xi xi2}}{Q}\:,
\end{equation}

\begin{equation}
\dot{p}_{\xi xij}=\left(\frac{p_{\xi xi(j-1)}^{2}}{Q}-k_{B}T\right)-p_{\xi xij}\frac{p_{\xi xi(j+1)}}{Q}\:,1<j<M\:,
\end{equation}

\begin{equation}
\dot{p}_{\xi xiM}=\left(\frac{p_{\xi xi(M-1)}^{2}}{Q}-k_{B}T\right)\:,
\end{equation}

\begin{equation}
\dot{\beta}_{j}=\frac{p_{\beta j}}{Q_{\beta}}\:,
\end{equation}

\begin{equation}
\dot{p}_{\beta1}=\left(\frac{p_{A}^{2}}{W}-k_{B}T\right)-p_{\beta1}\frac{p_{\beta2}}{Q_{\beta}}\:,
\end{equation}

\begin{equation}
\dot{p}_{\beta j}=\left(\frac{p_{\beta(j-1)}^{2}}{Q_{\beta}}-k_{B}T\right)-p_{\beta j}\frac{p_{\beta(j+1)}}{Q_{\beta}}\:,1<j<M\:,
\end{equation}

\begin{equation}
\dot{p}_{\beta M}=\left(\frac{p_{\beta(M-1)}^{2}}{Q_{\beta}}-k_{B}T\right)\:.
\end{equation}

The equations of motion for the $(y_{i},p_{yi})$
phase-space coordinates are identical to those given for $(x_{i},p_{xi})$
upon changing the subindex $x$ by $y$. $m$ is the carbon mass and
$W$ is the mass of the barostat, that is coupled to the in-plane
$(x_{i},p_{xi})$ and $(y_{i},p_{yi})$ phase-space coordinates. There
are $3NM$ thermostats, $(\xi_{xij}$,$\xi_{yij},\xi_{zij})$, with
mass $Q,$ and momentum $(p_{\xi xij},p_{\xi yij},p_{\xi zij})$,
with $j=1,\ldots,M$. Each of the $3N$ atomic momentum coordinates
is coupled to a different chain of $M$ Nosé-Hoover thermostats. The
equations of motion for $(\xi_{yij},p_{\xi yij})$ and $(\xi_{zij},p_{\xi zij})$
are identical to those given for $(\xi_{xij},p_{\xi xij})$ after
changing the subindex $x$ by $y$ or $z$, respectively. This massive
thermostatting of the system is mandatory for MD path-integral simulations
to avoid ergodicity problems. For classical MD simulations it would
be equally appropriate to use a unique thermostat chain for all the
atoms. The only reason for using the massive thermostatting here is
that the additional computational cost is low and thus the same home-made
computer code can be used for both classical MD simulations and quantun
PIMD simulations. The barostat is coupled to a chain of $M$ thermostats,
$\beta_{j},$ with mass $Q_{\beta}$, and momemtun $p_{\beta j}$.
The graphene simulations presented here were done with $M=4,$ $Q=2\times10^{7}$
eVfs$^{2}$, $Q_{\beta}=2\times10^{9}$ eVfs$^{2}$, and $W=3.2\times10^{15}$
eVfs$^{2}$. The internal in-plane stress is

\begin{equation}
\tau_{int}=\frac{1}{2A}\left[{\displaystyle \sum_{i=1}^{N}\left(\frac{p_{xi}^{2}+p_{yi}^{2}}{m}\right)}-(2A)\frac{d\phi}{dA}\right]
\end{equation}

The equations of motion were integrated by employing
explicit reversible integrators using factorization techniques for
the Liouville time evolution operator. We used the reversible reference
system propagator algorithm (RESPA), which allows to define different
time steps for the integration of fast and slow degrees of freedom.\citep{ma96}

\section{Spinodal relation\label{sec:Spinodal-relations}}

The stability condition of a 2D solid requires that the free energy
$F$ must be a convex function of its natural variables. In particular,
$\partial^{2}F/\partial A_{p}^{2}>0$. A mechanical instability appears
at the area $A_{p,C}$ if$\:$\citep{maris91,boronat94,herrero03}
\begin{equation}
\left(\frac{\partial^{2}F}{\partial A_{p}^{2}}\right)_{A_{p,C}}=0\:.\label{eq:deri_2=00003D0}
\end{equation}
The in-plane stress at the critical area is
\begin{equation}
\tau_{C}=-\left(\frac{\partial F}{\partial A_{p}}\right)_{A_{p,C}}\:.
\end{equation}
The Taylor expansion of $F$ at the critical area $A_{p,C}$ is, under
consideration of Eq. (\ref{eq:deri_2=00003D0}) and up to the order
$\mathcal{O}\left[(A_{p}-A_{p,C})^{4}\right]$, given as
\begin{equation}
F=F_{C}-\tau_{C}(A_{p}-A_{p,C})+a(A_{p}-A_{p,C})^{3}\:,
\end{equation}
where $F_{C}\equiv F(A_{p,C})$ and $a$ is a constant proportional
to the third $A_{p}-$derivative of $F$ at the critical area $A_{p,C}$.
The in-plane stress is obtained by the derivative of the last equation
as,
\begin{equation}
\tau=\tau_{C}-3a(A_{p}-A_{p,C})^{2}\:.\label{eq:tau}
\end{equation}
From this equation, 
\begin{equation}
A_{p}=A_{p,C}+\left(\frac{1}{3a}\right)^{1/2}\left(\tau_{C}-\tau\right)^{1/2}\label{eq:a_p}
\end{equation}
By considering the definition of the 2D compressional modulus in Eq.
(\ref{eq:B}) of Sec. \ref{sec:QHA-approach} and with the help of
Eqs. (\ref{eq:tau}) and (\ref{eq:a_p}), one gets,
\begin{equation}
B=(3a)^{1/2}A_{p,C}\left(\tau_{C}-\tau\right)^{1/2}+\mathcal{O}\left(\tau_{C}-\tau\right)\:,
\end{equation}
which gives us the dependence of the 2D compressional modulus $B$
with the in-plane stress $\tau$ expected close to the spinodal instability
of the layer at the critical stress $\tau_{C}$. Our simulation results
for $B(\tau)$ are in good agreement with the last equation.

\bibliographystyle{apsrev4-1}

\newpage{}

\end{document}